\documentclass[letterpaper, 10 pt, conference]{ieeeconf}  

\IEEEoverridecommandlockouts                              

\usepackage{amsmath} 
\usepackage{amssymb}  
\usepackage{color}
\usepackage{graphics}
\usepackage{graphicx}
\usepackage{epsfig}
\usepackage{epstopdf}

\usepackage{amsthm}

\usepackage[noadjust]{cite} 

\usepackage{tikz}
\usetikzlibrary{calc,positioning,shapes,shadows,arrows,fit}

\theoremstyle{plain}

\newtheorem{lemma}{Lemma}

\theoremstyle{definition}

\newtheorem{definition}{Definition}
\newtheorem{remark}{Remark}

\newtheorem{example}{Example}
\newtheorem{problem}{Problem}
 
\usepackage{algorithm}
\usepackage{algpseudocode}
\usepackage{pifont}

\newcommand{\T}{\mathcal{T}} 
\newcommand{\A}{\mathcal{A}} 

\newcommand{\init}{\mathit{init}}

\newcommand{\Nat}{\mathbb{N}} 

\newcommand{\AP}{\mathit{AP}} 

\renewcommand{\baselinestretch}{0.94}  

\title{\LARGE \bf
Cooperative Task Planning of Multi-Agent Systems Under Timed Temporal Specifications
}

\author{Alexandros Nikou, Jana Tumova and Dimos V. Dimarogonas
\thanks{The authors are with the ACCESS Linnaeus Center, School of Electrical
Engineering, KTH Royal Institute of Technology, SE-100 44, Stockholm,
Sweden and with the KTH Centre for Autonomous Systems. Email: {\tt\small \{anikou,tumova,dimos\}@kth.se}. This work was supported by the H2020 ERC Starting Grant BUCOPHSYS and the Swedish Research Council (VR).}
}

\begin{document}

\maketitle
\thispagestyle{empty}
\pagestyle{empty}

\begin{abstract}
In this paper the problem of cooperative task planning of multi-agent systems when timed constraints are imposed to the system is investigated. We consider timed constraints given by Metric Interval Temporal Logic (MITL). We propose a method for automatic control synthesis in a two-stage systematic procedure. With this method we guarantee that all the  agents  satisfy their own individual task specifications as well as that the team satisfies a team global task specification. 
\end{abstract}

\section{Introduction}

Cooperative control of multi-agent systems has traditionally focused on designing local control laws in order to achieve tasks such as consensus, formation, network connectivity, and collision avoidance (\cite{ren_beard_concensus, olfati_murray_concensus, jadbabaie_morse_coordination, zavlanos_connectivity, egerstedt_formation,  tanner_flocking, dimos_rendezvous_problem}). Over the last decade or so,  the field of control of multi-agent systems with complicated behavior under complex high-level task specifications has been gaining significant research attention. Such high-level tasks may have the form of ``Periodically survey regions A, B, C while avoiding region D", or ``Visit regions A, B, C, in this order", and many others. Multiple robotic vehicles then may perform these types of tasks faster and more efficiently than a single robot. In this work, we aim to introduce specific time bounds into the complex tasks, such as ``Periodically survey regions A, B, C, avoid region D and always keep the longest time between two consecutive visits to A below 5 time units", or ``Visit regions A, B, C, in this order within 10 time units".

The team of agents is usually associated with a set of tasks that should be fulfilled by the group of agents as a whole at the discrete level. A three-step hierarchical procedure to address such a problem is described as follows (\cite{vardi_2011_planning, fainekos_planning}): First the robot dynamics is abstracted into a finite or countable, discrete transition system using sampling or cell decomposition methods based on triangulations, rectangular or other partitions. Second, invoking ideas from verification methods, a discrete plan that meets the high-level task is synthesized. Third, the discrete plan is translated into a sequence of continuous controllers for the original system.

The specification language that has extensively been used to express the tasks is the Linear Temporal Logic (LTL) (see, e.g., \cite{loizou_2004}). LTL has proven a valuable tool for controller synthesis, because it provides a compact mathematical formalism for specifying desired behaviors of a system. There is a rich body of literature containing algorithms for verification and synthesis of system obeying temporal logic specifications (\cite{guo_2015_reconfiguration, frazzoli_vehicle_routing}). 
A common approach in multi-agent planning under LTL specifications is the consideration of a centralized, global task specification for the team of agents which is then decomposed into local tasks to be accomplished by the individual agents. 
For instance, the authors in \cite{belta_2010_product_system} utilized the parallel composition (synchronous products) of multi-robot systems in order to decompose a global specification that is given to a team of robots into individual specifications. This method has been proven computationally expensive due to state space explosion problem and in order to relax the computational burden the authors in \cite{belta_regular1, belta_regular2} proposed a method that does not require the computation of the parallel composition. This method, however, is restrictive to specifications that can be expressed in certain subclasses of LTL. In \cite{belta_cdc_reduced_communication}, the specification formula was given in LTL in parallel with the problem of minimum inter-robot communication.

Explicit time constraints in the system modeling have been included e.g., in \cite{belta_optimality}, where a method of automated planning of optimal paths of a group of agents satisfying a common high-level mission specification was proposed. The mission was given in LTL and the goal was the minimization of a cost function that captures the maximum time between successive satisfactions of the formula. Authors in \cite{quottrup_timed_automata, quadrup2} used a different approach, representing the motion of each agent in the environment with a timed automaton. The composition of the team automaton was achieved through synchronization and the UPPAAL verification tool (\cite{uppal}) was utilized for specifications given in Computational Tree Logic (CTL). In the same direction, authors in \cite{belta_2011_timed_automata} modeled the multi-robot framework with timed automata and weighted transition systems considering LTL specifications and then, an optimal motion of the robots satisfying instances of the optimizing proposition was proposed.

Most of the previous works on multi-agent planning consider temporal properties which essentially treat time in a qualitative manner. For real applications, a multi-agent team might be required to perform a specific task within a certain time bound, rather than at some arbitrary time in the future (quantitative manner). 
Timed specifications have been considered in \cite{liu_MTL, murray_2015_stl, baras_MTL, frazzoli_MTL}. In \cite{liu_MTL}, the authors addressed the problem of designing high-level planners to achieve tasks for switching dynamical systems under Metric Temporal Logic (MTL) specification and in \cite{murray_2015_stl}, the authors utilized a counterexample-guided synthesis to cyber-physical systems subject to Signal Temporal Logic (STL) specifications. In \cite{baras_MTL}, the MTL formula for a single agent was translated into linear constraints and a Mixed Integer Linear Programming (MILP) problem was solved. However, these works are restricted to single agent motion planning and are not expendable to multi-agent systems in a straightforward way. In \cite{frazzoli_MTL}, the vehicle routing problem was considered under the presence of MTL specifications. The approach does not rely on automata-based approach to verification, as it constructs a set of linear inequalities from MTL specification formula in order to solve an MILP problem. 

In this work, we aim at designing automated planning procedure for a team of agents that are given an individual, independent timed temporal specification each and a single global team specification. This constitutes the first step towards including time constraints to temporal logic-based multi-agent control synthesis. We consider a quantitative logic called Metric Interval Temporal Logic (MITL) (\cite{alur_mitl}) in order to specify explicit time constraints. The proposed solution is fully automated and completely desynchronized in the sense that a faster agent is not required to stay in a region and wait for the slower one. It is decentralized in  handling the individual specifications and centralized only in handling the global team specification. To the best of the authors' knowledge this is the first work that address the cooperative task planning for multi-agent systems under individual and global timed linear temporal logic specifications.

The remainder of the paper is structured as follows. In Sec. \ref{sec: preliminaries} a description of the necessary mathematical tools, the notations and the definitions are given. Sec. \ref{sec: prob_formulation} provides the model of the multi-agent system, the task specification, several motivation examples as well as the formal problem statement. Sec. \ref{sec: solution} discusses the technical details of the solution. Sec. \ref{sec: simulation_results} is devoted to an illustrative example. Finally, the conclusions and the future work directions are discussed in Sec. \ref{sec: conclusions}.
\vspace{-1mm}
\section{Notation and Preliminaries} \label{sec: preliminaries}

Given a set $S$, we denote by $|S|$ its cardinality and by $2^S$ the set of all its subsets. An infinite sequence of elements of $S$ is called a infinite word over the set $S$ and it is denoted by $w = w(0)w(1) \ldots $ The $i$-th element of a sequence is denoted with $w(i)$. We denote by $\mathbb{Q}_+, \mathbb{N}$ the set of positive rational and natural numbers including 0, respectively. Let us also define $\mathbb{T}_{\infty} = \mathbb{T} \cup \{\infty\}$ for a set of numbers $\mathbb{T}$.


\begin{definition} (\cite{alur1994}) A \emph{time sequence} $\tau = \tau(0) \tau(1) \cdots$ is a infinite sequence of time values $\tau(j) \in \mathbb{T} = \mathbb{Q}_{+}$, satisfying the following constraints:
\begin{itemize}
\item Monotonicity: 
$\tau(j) < \tau(j+1)$ for all $j \geq 0$. 
\item Progress: For every $t \in \mathbb{T}$, 
$\exists \ j \geq 1$, such that $\tau(j) > t$.
\end{itemize}
\end{definition}

An \emph{atomic proposition} $p$ is a statement over the problem variables and parameters that is either True $(\top)$ or False $(\bot)$ at a given time instance.

\begin{definition} (\cite{alur1994})
Let $\AP$ be a finite set of atomic propositions. A \emph{timed word} $w$ over the set $\AP$ is an infinite sequence $w = (w(0), \tau(0)) (w(1), \tau(1)) \cdots$ where $w(0) w(1) \ldots$ is an infinite word over the set $2^{\AP}$ and $\tau(0) \tau(1) \ldots$ is a time sequence with $\tau(j) \in \mathbb{T}, \ j \geq 0$.  A \emph{timed language} $\mathit{Lang}_\mathcal{T}$ over $AP$ is a set of timed words over $AP$.
\end{definition}

\subsection{Weighted Transition System}

\begin{definition}
A \emph{Weighted Transition System} (WTS) is a tuple $(S, S_0, \xrightarrow[~~]{}, d, AP, L)$ where
$S$ is a finite set of states;
$S_0 \subseteq S$ is a set of initial states;
$\xrightarrow[~~]{} \subseteq S \times S$ is a transition relation;
$d: \xrightarrow[~~]{} \xrightarrow[~~]{} \mathbb{T}$ is a map that assigns a positive weight to each transition;
$\AP$ is a finite set of atomic propositions; and
$L: S \xrightarrow[~~]{} 2^{AP}$ is a labeling function.
\end{definition}
\vspace{-2mm}
For simplicity, we use $s \rightarrow s'$ to denote the fact that $(s,s') \in \rightarrow$.

\begin{definition}
\label{run_of_WTS}
A \emph{timed run} of a WTS is an infinite sequence $r^t = (r(0), \tau(0))(r(1), \tau(1)) \ldots$,
such that 
$r(0) \in S_0$, and for all $j \geq 1$, $r(j) \in S$ and $r(j) \rightarrow r(j+1)$. The \emph{time stamps} $\tau_k(j), j \geq 0$ are inductively defined as
\begin{enumerate}
\item $\tau(0) = 0$.
\item $\displaystyle \tau(j+1) =  \tau(j) + d(r(j), r(j+1)), \ \forall \ j \geq 1.$
\end{enumerate}
\label{eq: timed_word_WTS}
Every timed run $r^t$ generates a \emph{timed word}
$w(r^t) = 
(L(r(0)), \tau(0)) \ (L(r(1)), \tau(1))\ldots$
over the set $2^{\AP}$ where $w(j) = L(r(j))$, $\forall \ j \geq 0$ is the subset of atomic propositions that are true at state $r(j)$ at time $\tau(j)$. 
\end{definition}

\subsection{Metric Interval Temporal Logic and Timed Automata}

The syntax of \emph{Metric Interval Temporal Logic (MITL)} over a set of atomic propositions $AP$ is defined by the grammar
\begin{equation} \label{eq: grammar}
\varphi := p \ | \ \neg \varphi \ | \ \varphi_1 \wedge \varphi_2 \ | \bigcirc_I \ \varphi \mid \Diamond_I \varphi \mid \square_I \varphi \mid  \varphi_1 \ \mathcal{U}_I \ \varphi_2
\end{equation}
where $p \in \AP$, and $\bigcirc$, $\Diamond$, $\square$ and $\mathcal U$ is the next, future, always and until temporal operator, respectively. $I \subseteq \mathbb{T}$ is a non-empty time interval in one of the following forms: $[i_1, i_2], [i_1, i_2),(i_1, i_2], $ $ (i_1, i_2), [i_1, \infty], (i_1, \infty)$ where $i_1, i_2 \in \mathbb{T}$ with $i_1 < i_2$. MITL can be interpreted either in continuous or point-wise semantics. We utilize the latter one and interpret MITL formulas over timed runs such as the ones produced by a WTS (Def.~\ref{run_of_WTS}).
\begin{definition} (\cite{pavithra_expressiveness}, \cite{quaknine_decidability})
Given a run $r^t = (r(0),\tau(0))(r(1),\tau(1)) \dots$ of a WTS and an MITL formula $\varphi$, we define $(r^t, i) \models \varphi$, for $\ i \geq 0$ (read $r^t$ satisfies $\varphi$ at position $i$) as follows
\begin{align*} \label{eq: for1}
(r^t, i) &\models p \Leftrightarrow p \in L(r(i)) \\
(r^t, i) &\models \neg \varphi \Leftrightarrow (r^t, i) \not \models \varphi \\
(r^t, i) &\models \varphi_1 \wedge \varphi_2 \Leftrightarrow (r^t, i) \models \varphi_1 \ \text{and} \ (r^t, i) \models \varphi_2 \\
(r^t, i) &\models \bigcirc_I \ \varphi \Leftrightarrow (r^t, i+1) \models \varphi \ \text{and} \    \tau(i+1) - \tau(i) \in I\\
(r^t, i) & \models \Diamond_I \varphi \Leftrightarrow \exists j, i \leq j, \ \text{s.t. } (r^t, j) \models \varphi, \tau(j)-\tau(i) \in {I} \\
(r^t, i) & \models \square_I \varphi \Leftrightarrow \forall j, i \leq j, \ \tau(j)-\tau(i) \in {I} \Rightarrow (r^t, j) \models \varphi  \\
(r^t, i) &\models \varphi_1 \ \mathcal{U}_I \ \varphi_2 \Leftrightarrow \exists j, i \leq j, \ \text{s.t. } (r^t, j) \models \varphi_2, \\ & \tau(j)-\tau(i) \in I \ \text{and } (r^t, k) \models \varphi_1 \ \text{for every} \ i \leq k < j 
\end{align*}

\end{definition}

\emph{Timed B\"uchi Automata (TBA)} were introduced in \cite{alur1994} and in this work, we also partially adopt the notation from \cite{bouyer_phd, tripakis_tba}. 
Let $X = \{x_1, x_2, \ldots, x_M\}$ be a finite set of \emph{clocks}. The set of \emph{clock constraints} $\Phi(X)$ is defined by the grammar
\begin{equation}
\phi :=  \top \mid \ \neg \phi \ | \ \phi_1 \wedge \phi_2 \ | \ x \bowtie c \ 
\end{equation}
where $x \in X$ is a clock, $c \in \mathbb{T}$ is a clock constant and $\bowtie \in  \{ <, >, \geq, \leq, = \}$. A clock \emph{valuation} is a function $\nu: X \rightarrow\mathbb{T}$ that assigns a real value to each clock. A clock $x_i$ has valuation $\nu_i$ for $i \in \{1, \ldots, M\}$, and $\nu = (\nu_1, \ldots, \nu_M)$. We denote by $\nu \models \phi$ the fact that the valuation $\nu$ satisfies the clock constraint $\phi$. 


\begin{definition}
A \emph{TBA} is a tuple $\mathcal{A} = (S, S^{\text{init}}, X, I,  
E, F, AP, \mathcal{L})$ where
$S$ is a finite set of locations;
$S^{\text{init}} \subseteq S$ is the set of initial locations;
$X$ is a finite set of clocks;
$I: S \rightarrow \Phi(X)$ is the invariant; 
$E \subseteq S \times \Phi(X) \times 2^X \times S$ gives the set of transitions;
$F \subseteq S$ is a set of accepting locations;
$\AP$ is a finite set of atomic propositions; and
$\mathcal{L}: S \rightarrow 2^{AP}$ labels every state with a subset atomic propositions.
\end{definition}


A state of $\mathcal{A}$ is a pair $(s, \nu)$ where $s \in S$ and $\nu$ satisfies the \emph{invariant} $I(s)$, i.e., $\nu \models I(s)$. The initial state of $\mathcal{A}$ is $(s(0), (0,\ldots,0))$, where $s(0) \in S_0$. Given two states $(s, \nu)$ and $(s', \nu')$ and an edge $e = (s, \gamma, R, s')$, there exists a \emph{discrete transition} $(s, \nu) \xrightarrow{e} (s', \nu')$ iff $\nu$ satisfies the \emph{guard} of the transition $\gamma$, i.e., $\nu \models \gamma$, $\nu' \models I(s')$, and $R$ is the \emph{reset set}, i.e., $\nu'_i = 0$ for $x_i \in R$ and $\nu'_i = \nu_i$ for $x_i \notin R$. Given a $\delta \in \mathbb{T}$, there exists a \emph{time transition} $(s, \nu) \xrightarrow{\delta} (s', \nu')$ iff $s = s', \nu' = \nu+\delta$ and $\nu' \models I(s)$. 
An infinite run of $\mathcal{A}$ starting at state $(s(0), \nu)$ is an infinite sequence of time and discrete transitions $(s(0), \nu(0))\xrightarrow{\delta_0} (s(0)', \nu(0)')\xrightarrow{e_0} (s(1), \nu(1)) \xrightarrow{\delta_1} (s(1)', \nu(1)') \ldots$, where $(s(0),\nu(0))$ is an initial state. 
This run produces the timed word $w = (\mathcal{L}(s(0)), \tau(0)) (\mathcal{L}(s(1)), \tau(1)) \ldots$ with $\tau(0) = 0$ and $\tau(i+1) = \tau(i) +\delta_i$,  $\forall \ i \geq 1$. The run is called \emph{accepting} if $s(i) \in F$ for infinitely many times. A timed word is \emph{accepted} if there exists an accepting run that produces it. 
The problem of deciding the emptiness of the language of a given TBA $\mathcal{A}$ is \emph{PSPACE}-complete \cite{alur1994}. In other words, we can synthesize an accepting run of a given a TBA $\mathcal{A}$, if one exists.

\begin{remark}
Traditionally, the clock constraints and the TBAs are defined with $\mathbb T = \Nat$, however, they can be extended to accommodate $\mathbb T = \mathbb Q_+ \cup \{0\}$. By multiplying all the rational numbers that are appearing in the state invariants and the edge constraints with their least common multiple, we have equivalently only natural numbers occurring to the TBA. For the sake of physical understanding of the timed properties of the under investigation framework, we will be working with $\mathbb{T} = \mathbb{Q}_{+} \cup \{0\}$.
\end{remark}

Any MITL formula $\varphi$ over $AP$ can be algorithmically translated to a TBA with the alphabet $2^\AP$, such that the language of timed words that satisfy $\varphi$ is the language of timed words produced by the TBA \cite{alur_mitl, maler_MITL_TA, MITL_ata}.



\vspace{-2mm}

\section{Problem Formulation} \label{sec: prob_formulation}

\subsection{System Model}

Consider a multi-agent team composed by $N$ agents operating in a bounded workspace $\mathcal{W}_0 \subseteq \mathbb{R}^n$. Let $\mathcal{I} = \left\{ 1,\ldots, N\right\}$ denote the index set of the agents. We assume that the workspace $\mathcal{W}_0$ is partitioned into a finite number (assume $W$) of regions of interest $\pi_1, \ldots, \pi_W$ where 
\begin{equation} \label{eq: partition}
\mathcal{W}_0 = \mathop{\bigcup} \limits_{i \in \mathcal{W}}^{} \pi_i \ \ \text{and} \ \ \pi_i \cap \pi_j \neq \emptyset , \forall \  i \neq j \ \ \text{with} \ \ i,j \in \mathcal{W}
\end{equation}
for the index set $\mathcal{W} = \{1,\ldots,W\}$. We denote by $\pi_i^k$ the agent $k$ being at region $\pi_i$, where $k \in \mathcal{I}, i \in \mathcal{W}$. In this work, we focus on interaction and high-level control strategies rather than on nonlinear models, and we assume that the dynamics of each agent is given by a single integrator
\begin{equation} \label{eq: system}
\dot{x}_i = u_i, \ i \in \mathcal{I}.
\end{equation}
The partitioned environment \eqref{eq: partition} is a discretization that allows us to control the agents with dynamics \eqref{eq: system} using finite models such as finite transition systems (e.g., \cite{fainekos_planning, tabuada_book_verification, girard_approximation, alur_pappas_abstractions}). We define a weighted transition system (see Def. \ref{def: transition_systems}) so that
\begin{itemize}
\item if there exists a controller $u_i, \ i \in \mathcal{I}$ such that the agent $k$ can be driven from any point within the region $\pi^i$ to a neighboring region $\pi^j$, then we allow for a transition $\pi_k^i \rightarrow_k \pi_k^j$ between the respective system states, and 
\item the weight of each transition {estimates} the time each agent needs in order to move from one region to another. In particular, the travel time is here determined as the worst-case shortest time needed to travel from an arbitrary point of the current region to the boundary of the following region. This estimate is indeed conservative, however, it is sufficient for specifications that we are generally interested in within multi-agent control. Namely, it is suitable for scenarios where tasks are given deadlines and upper rather than lower bound requirements are associated with events along the agents' runs.
\end{itemize}

\begin{definition} \label{def: transition_systems}
The motion of each agent $k \in \mathcal{I}$ in the workspace is modeled by a WTS $\mathcal{T}_k =(\Pi_k, \Pi_{k}^{\text{init}}, \rightarrow_k,d_k, AP_k, L_k)$ where
\begin{itemize}
\item $\Pi_k = \left\{ \pi_1^k, \pi_2^k, \ldots, \pi_W^k \right\}$ is the set of states of agent $k$. Any state of an agent $k$ can be denoted as $\pi_j^k \in \Pi_k$ for $k \in \mathcal{I}, j \in \mathcal{W}$. The number of states for each agent is $|\Pi_k| = W$.
\item $\Pi_k^{\text{init}} \subseteq \Pi_k$ is the initial states of agent $k$, i.e. the set of regions where agent $k$ may start.
\item $\rightarrow_k \subseteq \Pi_k \times \Pi_k$ is the transition relation. 
For example, by 
$\pi_3^3 \rightarrow_3 \pi_5^3$ we mean that the agent $3$ can move from region $\pi_3$ to region $\pi_5$.  
\item $d_k: \rightarrow_k \rightarrow \mathbb{T}$ is a map that assigns a positive weight (duration) to each transition. For example, $d_2(\pi_2^2, \pi_5^2) = 0.7, \ \text{where} \ \pi_2^2 \rightarrow_2 \pi_5^2$, means that agent $2$ needs at most $0.7$ time units to move from any point of region $\pi_2$ to the boundary of the neighboring region $\pi_5$.
\item $\AP_k$ is a finite set of atomic propositions known to agent $k$. Without loss of generality, we assume that $\AP_k \cap \AP_{k'} = \emptyset$ for all $k \neq k' \in \mathcal{I}$.
\item $L_k: \Pi_k \rightarrow 2^{\AP_k}$ is a labeling function that assigns to each state $\pi^k_j \in \Pi_k$ a subset of atomic propositions $AP_k$ that are satisfied when agent $k$ is in region $\pi_j$.
\end{itemize}
\end{definition}



\subsubsection{Individual Timed Runs and Words}
The behaviors of the individual agents can be captured through their timed runs and timed words. The timed run $r^t_k = (r_k(0), \tau_k(0))(r_k(1), \tau_k(1)) \cdots, \ k \in \mathcal{I}$ of each WTS $\mathcal{T}_k, \ k \in \mathcal{I}$ and the corresponding timed words $w(r_k^t) = (L_k(r_k(0)), \tau_k(0)) \ (L_k(r_k(1)), \tau_k(1)) \ \cdots$ are defined by using the terminology of Def. \ref{run_of_WTS}.

\smallskip

\subsubsection{Collective Timed Run and Word} \label{sec: collective_run} At the same time, the agents form a team and we are interested in their global, collective behaviors, which we formalize through the following definition.

\begin{definition} 
\label{def: collective_run}
Let $r_1^t, \ldots, r_N^t$ be  individual timed runs of the agents $1, \ldots, N$, respectively, 
as defined above. Then, the \emph{collective timed run} $r_G = (r_G(0), \tau_G(0)) (r_G(1), \tau_G(1)) \ldots$ of the team of agents is defined inductively as follows
\begin{enumerate}
\item $(r_G(0), \tau_G(0)) = ( (r_1(0), \ldots, r_N(0)) , \tau_G(0))$.
\item Let $(r_G(i), \tau_G(i)) = ((r_1(i_1), \ldots, r_N(i_N)) , \tau_G(i))$, where $i \geq 0$ be the current state and time stamp of the collective timed run. Then 
the next state and time stamp $(r_G(i+1), \tau_G(i+1)) = ((r_1(j_1), \ldots, r_N(j_N)) , \tau_G(i+1))$ are given by the following
\begin{itemize}
\item $\ell = \underset{k \in \mathcal{I}}{\text{argmin}}\{ \tau_k(i_k+1)\}$.
\item $\tau_G(i+1) = \tau_\ell(i_\ell+1)$.
\item $r_k(j_k) =
\begin{cases}
r_\ell(i_\ell+1) & \ \text{if} \ k = \ell \\
r_k(i_\ell) & \ \text{if} \ k \neq \ell.
\end{cases} $
\end{itemize}
\end{enumerate}
Intuitively, given the current states $r_1(i_1),\ldots,r_N(i_N)$ and the next states $r_1(i_1+1),\ldots,r_N(i_N+1)$ of the individual agents at time $\tau_G(i)$, $\ell$ is the index of the agent $k$ who will finish its current transition from $r_\ell(i_\ell)$ to $r_\ell(i_\ell + 1)$ the soonest amongst all. The time of agent $\ell$'s arrival to its next state $r_\ell(i_\ell + 1)$ becomes the new time stamp $\tau_G(i+1)$ of the collective timed run. The next state of the collective timed run reflects that each agent $k$ which cannot complete its transition from $r_k(i_k)$ to $r_k(i_k+1)$ before $\tau_G(i+1)$ remains in $r_k(i_k)$.

\end{definition}
\vspace{-3mm}
In what follows, 
$r_G^t = (r_G(0), \tau_G(0)) (r_G(1), \tau_G(1)) \ldots,$
where $r_G(i) = (r_1(i_1), \ldots, r_N(i_N)), \ i, i_k \geq 0$ and $k \in \mathcal{I}$ denotes the collective timed run. 
\vspace{-3mm}
\begin{definition}
We define the global set of atomic propositions $\AP_G = \displaystyle \bigcup_{k=1}^{N} \AP_k$ and for every state $r_G(i) = (r_1(i_1), \ldots, r_N(i_N))$ of a collective timed run, where $i, i_k \geq 0$ and $k \in \mathcal{I}$, we define the labeling function $L_G:\Pi_1 \ldots \Pi_N \rightarrow \AP_G$ as $L_G(r_G(i)) =  \bigcup_{k=1}^{N} L_k(r_k(i_k))$.
\end{definition}
A collective timed run $r_G^t$ thus naturally produces a timed word $w_G^t = (L_G(r_G(0)), \tau_G(0)) (L_G(r_G(1)), \tau_G(1)) \ldots$ over $\AP_G$.
\vspace{-2mm}
\begin{example} \label{example: team_run}
Consider $N=2$ robots operating in a workspace with $\mathcal{W} = \pi_1 \cup \pi_2 \cup \pi_3, W_0 = 3$ and $\mathcal{I} = \{1,2\}$  modeled as the WTSs illustrated in Fig. \ref{fig: trans_system_example}. Let $\AP_1 = \{\mathit{green}\}$, and $\AP_2 = \{\mathit{red}\}$. The labeling functions are $L_1(\pi_1^1) = \{green\}, L_1(\pi_2^1) = L_1(\pi_3^1) = \emptyset$, and $L_2(\pi_1^2) = L_2(\pi_2^2) = \emptyset, L_2(\pi_3^2) = \{red\}$.
\vspace{-7mm}
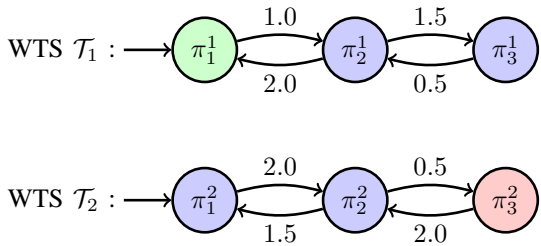
\begin{figure}[ht!]
\centering
\begin{tikzpicture}[scale = 1.0]       
\node(pseudo1) at (-1.2,0){};
\node(0) [line width = 1.0] at (0,0)[shape=circle,draw][fill=green!20]           {$\pi_1^1$};
\node(1) [line width = 1.0] at (2,0)[shape=circle,draw][fill=blue!20]         {$\pi_2^1$};
\node(5) [line width = 1.0] at (4,0)[shape=circle,draw][fill=blue!20]         {$\pi_3^1$};

\node(pseudo2) at (-1.2,-2.0){};
\node(2) [line width = 1.0] at (0,-2.0)[shape=circle,draw][fill=blue!20]           {$\pi_1^2$};
\node(3) [line width = 1.0] at (2,-2.0)[shape=circle,draw][fill=blue!20]         {$\pi_2^2$};
\node(6) [line width = 1.0] at (4,-2.0)[shape=circle,draw][fill=red!20]         {$\pi_3^2$};

\path [->] [line width = 1.0]
  (0)     edge       [bend left = 15]              node  [above]  {$1.0$}  (1)
  (1)     edge     [bend right = -15]           node  [below]  {$2.0$}     (0)
  (1)     edge     [bend right = -15]           node  [above]  {$1.5$}     (5)
    (5)     edge     [bend right = -15]           node  [below]  {$0.5$}     (1)

  (2)     edge       [bend left = 15]              node  [above]  {$2.0$}  (3)
  (3)     edge     [bend right = -15]           node  [below]  {$1.5$}     (2)
   (3)     edge     [bend right = -15]           node  [above]  {$0.5$}     (6)
      (6)     edge     [bend right = -15]           node  [below]  {$2.0$}     (3)

  (pseudo1) edge                                       (0)
  (pseudo2) edge                                       (2);

\node at (-1.9, 0.0) {WTS $\T_1$ :};
\node at (-1.9, -2.0) {WTS $\T_2$ :};
\end{tikzpicture}
\caption{WTSs $T_1, T_2$ representing two agents in $\mathcal W$. $\Pi_1 = \{\pi_1^1, \pi_2^1, \pi_3^1\}$, $\Pi^{\text{init}}_1 = \{\pi_1^1\}$, $\Pi_2 = \{\pi_1^2, \pi_2^2, \pi_3^2\}, \Pi^{\text{init}}_2 = \{\pi_1^2\}$, the transitions are depicted as arrows which are annotated with the corresponding weights.
}
\label{fig: trans_system_example}
\end{figure}

\noindent Examples of the agents' runs are:
\begin{align}
r_1^t  = & (r_1(0) = \pi_1^1, \tau_1(0) = 0.0)(r_1(1) = \pi_2^1, \tau_1(1) = 1.0) \notag \\
&(r_1(2) = \pi_3^1, \tau_1(2) = 2.5) (r_1(3) = \pi_2^1, \tau_1(3) = 3.0) \notag \\
& (r_1(4) = \pi_1^1, \tau_1(4) = 5.0) 
\ldots \notag \\
r_2^t = & (r_2(0) = \pi_1^2, \tau_2(0) = 0.0)(r_2(1) = \pi_2^2, \tau_2(1) = 2.0) \notag \\
& (r_2(2) = \pi_3^2, \tau_2(2) = 2.5) (r_2(3) = \pi_2^2, \tau_2(3) = 4.5) \notag \\
& (r_2(4) = \pi_3^2, \tau_2(4) = 5.0) \ldots \notag 
\end{align}

Given $r_1^t$ and $r_2^t$ the collective run $r_G$ is given according to Def. \ref{def: collective_run} as follows:
\begin{align}
r_G^t = & (\underbrace{(\pi_1^1, \pi_1^2)}_{r_G(0)}, \tau_G(0) = 0.0)(\underbrace{(\pi_2^1, \pi_1^2)}_{r_G(1)}, \tau_G(1) = 1.0) \notag \\
& (\underbrace{(\pi_2^1, \pi_2^2)}_{r_G(2)}, \tau_G(2) = 2.0) (\underbrace{(\pi_3^1, \pi_3^2)}_{r_G(3)}, \tau_G(3) = 2.5) \notag \\
& (\underbrace{(\pi_2^1, \pi_3^2)}_{r_G(4)}, \tau_G(4) = 3.0) 
(\underbrace{(\pi_2^1, \pi_2^2)}_{r_G(5)}, \tau_G(5) = 4.5) \notag \\
&(\underbrace{(\pi_1^1, \pi_3^2)}_{r_G(6)}, \tau_G(6) = 5.0) \ldots 
\notag
\end{align}

The produced collective timed word is 
\begin{align*}
w_G^t= & (\{green\},0.0)(\emptyset,1.0)(\emptyset,2.0)  (\{red\},2.5) \\ &(\{red\},3.0)(\emptyset,4.5)(\{green,red\},5.0)\ldots.
\end{align*}


\end{example}

\subsection{Specification}
Several different logics have been designed to express timed properties of real-time systems, such as MTL \cite{koymans_MTL} that extends the until operator of LTL with a time interval.
Here, we consider a fragment of MTL, called MITL (see Sec. \ref{sec: preliminaries} for definition) which has been proposed in \cite{alur_mitl}. Namely, we utilize its point-wise semantics and interpret its formulas over timed runs. Unlike MTL, MITL excludes punctual constraints on the until operator. 
For instance, the formula $\square (a \Rightarrow \Diamond_{=1}b)$ 
saying that every $a$ is followed by a $b$ precisely 1 time unit later,  is not allowed in MITL, 
whereas $\square(a \Rightarrow \Diamond_{(0, 1]} b)$, saying that every $a$ is followed by a $b$ at most after 1 time unit later, is. While MTL formulas cannot be generally translated into TBAs, MITL formulas can \cite{alur_mitl}.

\smallskip

\subsubsection{Local Agent's Specification}

Each agent $k$, $k \in \mathcal{I}$ is given an individual, local, independent specification in the form of a MITL formula $\varphi_k$ over the set of atomic propositions $\AP_k$. The satisfaction of $\varphi_k$ is decided from the agent's own perspective, i.e., on the timed run $r_k^t$.

\smallskip

\subsubsection{Global Team Specification} \label{sec: spec_satisf}

In addition, the team of agents is given a global team specification, which is a MITL formula $\varphi_G$ over the set of atomic propositions $\AP_G$. The team specification satisfaction is decided on the collective timed run~$r_G^t$.

\addtocounter{example}{-1}
\begin{example}[Continued]
Recall the two agents from Example \ref{example: team_run}. Each of the agents is given a local, independent, specification and at the same time, the team is given an overall goal that may require collaboration or coordination. Examples of local specification formulas are $\varphi_1 = \square \Diamond_{\leq 10} (\mathit{green})$ and $\varphi_2 = \square( \mathit{red} \Rightarrow \bigcirc \square_{\leq 5} (\neg \mathit{red}))$ stating that ``The green region is periodically visited with at most 10 time units between two consecutive visits'' and ``Whenever a red region is visited, it will not be visited for the following 5 time units again'', respectively. While $\varphi_1$ is satisfied on $r_1^t$, $\varphi_2$ is not satisfied on $r_2^t$. An example of the global specification is $\varphi_G = \square \Diamond_{\leq 5}  (\mathit{green} \wedge \mathit{red})$ that imposes requirement on the agents' collaboration; it states that agents 1 and 2 will periodically simultaneously visit the green and the red region, respectively, with at most 5 time units between two consecutive visits.
\end{example}

\subsection{Problem Statement}

\begin{problem}[Run Synthesis] \label{problem: basic_prob}
Given $N$ agents governed by dynamics as in \eqref{eq: system}, a task specification MITL formula $\varphi_G$ for the team of robots, over a set of atomic propositions $AP_G$ and $N$ local task specifications $\varphi_k$ over $AP_k, \ k \in \mathcal{I}$, synthesize a sequence of individual timed runs $r_1^t, \ldots, r_N^t$ such that the following hold
\begin{equation} \label{eq: problem_adf}
\left(r_G^t \models \varphi_G \right) \wedge \left( r_1^t \models \varphi_1 \wedge \ldots \wedge r_N^t \models \varphi_N \right).
\end{equation}
\end{problem}

Though it might seem that the satisfaction of the individual specifications $\varphi_1,\ldots,\varphi_N$ can be treated as the satisfaction of the formula $\bigwedge_{k \in \mathcal{I}}{\varphi_k}$ on the collective timed run $r_G^t$, this is generally not the case, as demonstrated through the following example:

\addtocounter{example}{-1}
\begin{example}[Continued]
Recall the two agents from Example \ref{example: team_run} and a local specification $\varphi_2 = \square( \mathit{red} \Rightarrow \bigcirc \square_{\leq 2} (\neg \mathit{red}))$. While this specification is satisfied on $r_2^t$ since $w(r_2^t) =  (\emptyset,0.0)(\emptyset,2.0) (\{red\},2.5)(\emptyset,4.5)(\{green,red\},5.0) \ldots$, it can be easily seen that it is not satisfied on $r_G^t$.
\end{example}
Formally, we have
\begin{equation}
r_G^t \models  \bigwedge_{k \in \mathcal{I}}{\varphi_k} \nLeftrightarrow r_1^t \models \varphi_1 \wedge \ldots \wedge r_N^t \models \varphi_N. \label{eq: remark_import}
\end{equation}
Hence, Problem~\ref{problem: basic_prob} may not be treated in a straightforward, fully centralized way. We propose a two-stage solution that first pre-computes all timed runs of the individual agents in a decentralized way and stores them efficiently in weighted transition systems enhanced with a B\"uchi acceptance condition. Second, these are combined and inspected with respect to guaranteeing the satisfaction of the team specification by the collective timed run. 

\section{Proposed Solution} \label{sec: solution}

In this section, we introduce a systematic solution to Problem~\ref{problem: basic_prob}.
Our overall approach builds on the following steps:
\begin{enumerate}
\item We construct TBAs $\mathcal{A}_k, \ k \in \mathcal{I}$ and $\mathcal{A}_G$ that accept all the timed words satisfying the specification formulas $\varphi_k, \ k \in \mathcal{I}$ and $\varphi_G$, respectively (Sec. \ref{sec: 4a}).
\item We construct a \emph{local B\"uchi WTS} $\widetilde{\mathcal{T}}_k = \mathcal{T}_k \otimes \mathcal{A}_k$, for all $\ k \in \mathcal{I}$. The accepting timed runs of $\widetilde{\T}_k$ are the timed runs of the $\mathcal{T}_k$ that satisfy the corresponding local specification formula $\varphi_k, k \in \mathcal{I}$ (Sec. \ref{sec: 4b}).
\item We construct a  \emph{product B\"uchi WTS} $\mathcal{T}_G = \widetilde{\mathcal{T}}_1 \otimes \cdots \otimes \widetilde{\mathcal{T}}_N$ such that its timed runs are collective timed runs of the team and their projections onto the agents' individual timed runs are admissible by the local B\"uchi WTSs $\widetilde \T_1, \ldots \widetilde \T_N$ respectively (Sec. \ref{sec: 4c}).
\item We construct a \emph{global B\"uchi WTS} $\widetilde{\mathcal{T}}_G = \mathcal{T}_G \otimes \mathcal{A}_G$. The accepting timed runs of the $\widetilde{\mathcal{T}}_G$ are the timed runs of the $\T_G$ that satisfy the team formula $\varphi_G$ (Sect. \ref{sec: product_buchi_tba}).
\item We find an accepting timed run $\widetilde{r}_G^t$ of the global B\"uchi WTS $\widetilde{\mathcal{T}_G}$ and project it onto timed runs of the product B\"uchi WTS $\T_G$, then onto timed runs of the local B\"uchi WTSs $\widetilde \T_1,\ldots, \widetilde \T_N$, and finally onto individual timed runs $r_1^t, \ldots, r_N^t$ of the original WTSs $\T_1,\ldots, \T_N$. By construction, $r_1^t, \ldots, r_N^t$ are guaranteed to satisfy $\varphi_1,\ldots, \varphi_N$, respectively, and furthermore $r_G^t$ satisfies $\varphi_G$ (Sec. \ref{sec: projection}).

\end{enumerate}

{
\subsection{Construction of TBAs} \label{sec: 4a}

As stated in Sec. \ref{sec: preliminaries}, every MITL formula $\varphi$ can be translated into a language equivalent TBA. Several approaches are proposed for that purpose, for instance \cite{maler_MITL_TA, alur_mitl, nickovic_timed_aut, MITL_ata}. Here, we translate each local specification $\varphi_k$, where $k \in \mathcal I$ into a TBA $\mathcal{A}_k = (S_k, S^\text{init}_k, X_k, I_k, E_k, \mathcal F_k, AP_k, \mathcal{L}_k)$, and the global specification $\varphi_G$ into a TBA $\A_G = (S_G,  S^\text{init}_G, X_G, I_G, E_G, \mathcal F_G, AP_G, \mathcal{L}_G)$.

}

\subsection{Construction of the local B\"uchi WTSs $\widetilde \T_1,\ldots,\widetilde \T_N$} \label{sec: 4b}
\begin{definition}
Given a WTS $\mathcal{T}_k =(\Pi_k, \Pi_{k}^{\text{init}}, \rightarrow_k, AP_k, L_k, d_k)$, and a TBA $\A_k = (S_k,  S^\text{init}_k, X_k, I_k, E_k, F_k, AP_k, \mathcal{L}_k)$ with $M_k = |X_k|$ and $C^{\mathit{max}}_k$ being the largest constant appearing in $\A_k$, we define their \emph{local B\"uchi WTS} $\widetilde{\T}_k = \mathcal{T}_k \otimes \A_k = (Q_k, Q_{k}^{\init}, {\rightsquigarrow}_{k}, \widetilde{d}_k, \widetilde{F}_k, AP_k, \widetilde{L}_k)$ as follows:
\begin{itemize}
  \item {$Q_k \subseteq \{(r_k,s_k) \in \Pi_k \times S_k : L_k(r_k) = \mathcal{L}_k(s_k)\} \times \mathbb{T}_\infty^{M_k} $.}
  \item $Q_{k}^{\init} = \Pi_k^{\init} \times S_k^{\init} \times \underbrace{\{0\} \times \ldots \times \{0\}}_{M_k \ products}$.
  \item $q \, {\rightsquigarrow}_k \, q'$ iff
  \begin{itemize}
    \item[$\circ$] $q = (r,s,\nu_1,\ldots,\nu_{M_k}) \in Q_k$, \\ $q' = (r',s',\nu_1',\ldots,\nu_{M_k}') \in Q_k$,
    \item[$\circ$] $r \, \rightarrow_k r'$, and
    \item[$\circ$] there exists $\gamma, R$, such that $(s,\gamma,R,s') \in E_k$, $\nu_1,\ldots,\nu_{M_k} \models \gamma$, $\nu_1',\ldots,\nu_{M_k}' \models I_k(s')$, and for all $i\in \{1,\ldots, M_k\}$
    \begin{equation*}
    \nu_i' = 
    \begin{cases}
        0,      & \text{if } x_i \in R \\
        \nu_i + d_k(r, r'), &  \text{if }  x_i \not \in R \text{ and } \\ & \nu_i + d_k(r, r') \leq C^{\mathit{max}}_k \\
        \infty, & \text{otherwise}.
      \end{cases}
    \end{equation*}
  \end{itemize}
    Then $\widetilde{d}_k(q,q') = d_k(r,r')$.
    \item $\widetilde{F}_k = \{(r_k,s_k,\nu_1,\ldots,\nu_{M_k}) \in Q_k : s_k \in F_k\}$.
    \item $\widetilde{L}_k(r_k, s_k, \nu_1, \ldots, \nu_{M_k}) = L_k(r_k)$.
\end{itemize}
\label{def:localBWTS}
\end{definition}

Each local B\"uchi WTS $\widetilde \T_k, k \in \mathcal I$ is in fact a WTS with a B\"uchi acceptance condition $\widetilde{F}_k$. A timed run of $\widetilde \T_k$ can be written as $\widetilde{r}_k^t = (q_k(0), \tau_k(0))(q_k(1), \tau_k(1)) \ldots$ using the terminology of Def. \ref{run_of_WTS}. It is \emph{accepting} if $q_k(i) \in \widetilde F_k$ for infinitely many $i \geq 0$.
An accepting timed run of  $\widetilde{\T}_k$ projects onto a timed run of $\T_k$ that satisfies the local specification formula $\varphi_k$ by construction. Formally, the following lemma, whose proof follows directly from the construction and and the principles of automata-based LTL model checking (see, e.g., \cite{katoen}), holds:
\vspace{-2mm}
\begin{lemma} \label{eq: lemma_1}
Consider an accepting timed run $\widetilde{r}_k^t = (q_k(0), \tau_k(0))(q_k(1), \tau_k(1)) \ldots$ of the local B\"uchi WTS $\widetilde \T_k$ defined above, where $q_k(i) = (r_k(i), s_k(i), \nu_{k, 1}, \ldots, \nu_{k, M_k})$ denotes a state of $\mathcal{\widetilde T}_k$, for all $i \geq 1$. 
The timed run $\widetilde{r}_k^t$ projects onto the timed run $r_k^t = (r_k(0), \tau_k(0))(r_k(1), \tau_k(1)) \ldots $ of the WTS $\mathcal{T}_k$ that produces the timed word $w(r_k^t) = (L_k(r_k(0)), \tau_k(0))(L_k(r_k(1)), \tau_k(1)) \ldots $ accepted by the TBA $\mathcal{A}_k$ via its run $\rho_k = s_k(0)s_k(1) \ldots$ Vice versa, if there exists a timed run $r_k^t = (r_k(0),\tau_k(0))(r_k(1),\tau_k(1))\ldots$ of the WTS $\T_k$ that produces a timed word $w(r_k^t) = (L_k(r_k(0)), \tau_k(0))(L_k(r_k(1)), \tau_k(1)) \ldots$ accepted by the TBA $\A_k$ via its run $\rho_k = s_k(0)s_k(1)\ldots$ then there exist the accepting timed run $\widetilde{r}_k^t = (q_k(0),\tau_k(0))(q_k(1),\tau_k(1)) \ldots$ of $\widetilde{\T}_k$, where $q_k(i)$ denotes $(r_k(i),s_k(i),\nu_{k,1}(i), \ldots, \nu_{k,M_k}(i))$ in $\widetilde{\T}_k$. 
\end{lemma}










\subsection{Construction of the product B\"uchi WTS $\mathcal{T}_G$} \label{sec: 4c}

Now we aim to construct a finite product WTS $\mathcal{T}_G$ whose timed runs represent the collective behaviors of the team and whose B\"uchi acceptance condition ensures that the accepting timed runs account for the local specifications. In other words, $\mathcal{T}_G$ is a product of all the local WTS $\widetilde{\T}_k$ built above.
In the construction of $\mathcal{T}_G$, we need to specifically handle the cases when transitions of different agents are associated with different time durations, i.e, different transition weights. To this end, we introduce a vector $b = (b_1, \ldots, b_N) \in \mathbb{T}^N$. Each element of the vector is a rational number $b_k \in \mathbb{T}, k \in \mathcal{I}$ which can be either $0$, when the agent $k$ has just completed its transition, or the time elapsed from the beginning of the agent's current transition, if this transition is not completed, yet. The state of the team of agents is then in the form $q_G = (q_1, \ldots, q_N, b_1, \ldots, b_N, \ell)$ where $q_k$ is a state of $\widetilde \T_k$, for all $k \in \mathcal{I}$, and $\ell \in \mathcal{I}$ has a special meaning in relation to the acceptance condition of $\T_G$ that will become clear shortly. Taking the above into consideration we define the global model $\T_G$ as follows:
\begin{definition}
Given $N$ local B\"uchi WTSs $\widetilde{\T}_1,\ldots,\widetilde{\T}_N$ from Def.~\ref{def:localBWTS}, their \emph{product B\"uchi WTS} $\mathcal{T}_G = \widetilde{\T}_1 \otimes \ldots \otimes \widetilde{\T}_N =(Q_G, Q_{G}^{\init}, \rightarrow_G, d_G, F_G, \AP_G, L_G)$ is defined as follows:
\begin{itemize}
\item {$Q_G \subseteq Q_1 \times \cdots \times Q_N \times \mathbb{T}^N \times \{1, \ldots, N\}$.}
\item $Q_{G}^{\init} = Q_{1}^{\init} \times \ldots \times Q_{N}^{\init} \times \underbrace{\{0\} \times \ldots \times \{0\}}_{N \ products} \times \{1\}$.
\item $q_G \rightarrow_G q_G'$ iff
\begin{itemize}
\item[$\circ$] $q_G =(q_1, \ldots, q_N, b_1, \ldots, b_N,\ell) \in Q_G, \\
q_G' = (q'_1, \ldots, q'_N, b'_1, \ldots, b'_N,\ell') \in Q_G$,
\item[$\circ$] {$\exists \ q''_k \in Q_k : \, q_k {\rightsquigarrow}_k \, q''_k$, for some $k \in \mathcal{I}$},

\item[$\circ$] \[b_k' =
  \begin{cases}
    0, & \text{if } b_k + d_{\mathit{min}} = \widetilde{d}_k(q_k,q_k'') \\ &\text{and } q_k' = q_k''  \\
    b_k + d_{\mathit{min}}, &\text{if } b_k + d_{\mathit{min}} < \widetilde{d}_k(q_k,q_k'') \\ &\text{and } q_k' = q_k 
  \end{cases}
\]  
where $d_{\mathit{min}} = \underset{k\in \{1,\ldots,N\}}{\text{min}} (\widetilde{d}_k(q_k,q_k'') - b_k)$ is (loosely speaking) the smallest time step that can be applied, and

\item[$\circ$] \[\ell' =
  \begin{cases}
    \ell,      & \text{if } q_\ell \not \in \widetilde{F}_\ell \\
    ((\ell + 1) \mod N), &  \text{otherwise}
  \end{cases}
\]  
\end{itemize}
Then $d_G(q_G,q_G') = d_{\mathit{min}}$.
\item $F_G = \{(q_1, \ldots, q_N, b_1, \ldots, b_N, N) \in Q_G : q_N \in \widetilde{F}_N\}$.
\item $AP_G = \displaystyle \bigcup_{k=1}^{N} AP_k$.
\item $L_G((q_1,\ldots,q_N,b_1,\ldots, b_N,\ell) = \displaystyle \bigcup_{k=1}^{N} \widetilde{L}_k(q_k)$.
\end{itemize}
\end{definition}

The product WTS $\T_G$ is again a WTS with a B\"uchi acceptance condition. Informally, the index $\ell$ in a state $q_G =(q_1, \ldots, q_N, b_1, \ldots, b_N,\ell) \in Q_G$ allows to project an accepting timed run of $\T_G$ onto an accepting run of every one of the local B\"uchi WTS. The construction is based on the standard definition of B\"uchi automata intersection (see, e.g.,~\cite{katoen}).

The following lemma follows directly from the construction and and the principles of automata-based LTL model checking (see, e.g., \cite{katoen}):
\vspace{-1mm}
\begin{lemma} \label{eq: lemma_2}
For all $k \in \mathcal I$, an accepting timed run ${r}_G^t$ of the product B\"uchi WTS ${\mathcal{T}}_G$ 
projects onto an accepting timed run $r_k^t$
of the local B\"uchi WTS $\widetilde\T_k$ that produces a timed word 
$w(r_k^t)$ 
accepted by the corresponding TBA $\mathcal{A}_k$. 
Vice versa, if there exists a timed run $r_k^t$ 
of the local B\"uchi WTS $\widetilde\T_k$ that produces a timed word $w(r_k^t)$ 
accepted by the TBA $\A_k$ for each $k \in \mathcal I$, 
then there exist an accepting timed run ${r}_G^t$ 
of ${\T}_G$. 
\end{lemma}

\vspace{-2mm}
\subsection{Construction of the global B\"uchi WTS $\widetilde{\mathcal{T}}_G$} \label{sec: product_buchi_tba}

\begin{definition}
Finally, given the product B\"uchi WTS $\mathcal{T}_G =(Q_G, Q_{G}^{\text{init}}, \rightarrow_G, d_G, F_G, AP_G, L_G)$, and a TBA $\A_G = (S_G,  S^\text{init}_G, X_G, I_G, E_G, \mathcal F_G, AP_G, \mathcal{L}_G)$ that corresponds to the team specification formula $\varphi_G$ with $M_G = |X_G|$ and $C^{\mathit{max}}_G$ being the largest constant appearing in $\A_G$, we define their product WTS $\widetilde{\T}_G = \mathcal{T}_G \otimes \A_G = (\widetilde{Q}_G, \widetilde{Q}_{G}^{\init}, \rightsquigarrow_{G},$ $\widetilde{d}_G, \widetilde{F}_G, AP_G, \widetilde{L}_G)$ as follows:
\begin{itemize}
  \item $\widetilde{Q}_G \subseteq \{(q,s) \in Q_G \times S_G : L_G(q) = \mathcal{L}_G(s)\} \times \mathbb{T}_\infty^{M_G}$.
  \item $\widetilde{Q}_{G}^{\init} = Q_G^{\init} \times S_G^{\init} \times \underbrace{\{0\} \times \ldots \times \{0\}}_{M_G - products} \times \{1,2\} $.
  \item $q \rightsquigarrow_G q'$ iff
  \begin{itemize}
    \item[$\circ$] $q = (r,s,\nu_1,\ldots,\nu_{M_G}, \ell) \in Q_G$ , \\ $q' = (r',s',\nu_1',\ldots,\nu_{M_G}',\ell') \in Q_G$,
    \item[$\circ$] $r \rightarrow_G r'$, and
    \item[$\circ$] there exists $\gamma, R$, such that $(s,\gamma,R,s') \in E_G$, $\nu_1,\ldots,\nu_{M_G} \models \gamma$, $\nu_1', \ldots, \nu_{M_G}' \models I_G(s')$, and for all $i\in \{1,\ldots, M_G\}$
    \begin{equation*}
    \nu_i' = 
    \begin{cases}
        0,      & \text{if } x_i \in R \\
        \nu_i + d_G(r, r'), &  \text{if }  x_i \not \in R \text{ and } \\ & \nu_i + d_G(r, r') \leq C^{\mathit{max}}_G \\
        \infty, & \text{otherwise}
      \end{cases}
    \end{equation*}
    \item[$\circ$] \[\ell' =
  \begin{cases}
    1      \text{ if } \ell = 1 \text { and } r \not \in {F}_G, \text{ or } \ell = 2 \text{ and } s \in \mathcal F_G \\
    2   \text{ otherwise}
  \end{cases}
\]  
  \end{itemize}
    Then $\widetilde{d}_G(q,q') = d_G(r,r')$.
    \item {$\widetilde{F}_G = \{(r,s,\nu_1,\ldots,\nu_{M_G},1) \in Q_G : r \in  F_G\}$.}
    \item $\widetilde{L}_G(r_G, s_G, \nu_1, \ldots, \nu_{M_G}) = L_G(r_G)$.
\end{itemize}
\end{definition}

Analogously to above, the global B\"uchi WTS $\widetilde \T_G$ is a WTS with a B\"uchi acceptance condition. An accepting timed run of $\widetilde{T}_G$ guarantees the satisfaction of the team specification formula $\varphi_G$ by construction. Furthermore, the projected individual timed runs of the original $\T_1, \ldots, \T_N$ satisfy their respective local specifications. The following lemma follows directly from the construction and and the principles of automata-based LTL model checking (see, e.g., \cite{katoen}):
\vspace{-1mm}
\begin{lemma} \label{eq: lemma_3}
An accepting timed run $\widetilde{r}_G^t$ of the global B\"uchi WTS $\widetilde{\mathcal{T}}_G$ 
projects onto an accepting timed run $r_G^t$ 
of the product B\"uchi WTS $\mathcal{T}_G$ that produces a timed word 
$w(r_G^t)$ 
accepted by the TBA $\mathcal{A}_G$. 
Vice versa, if there exists a timed run $r_G^t$ 
of the product B\"uchi WTS $\mathcal{T}_G$ that produces a timed word $w(r_G^t)$ 
accepted by the TBA $\A_G$ 
then there exist an accepting timed run $\widetilde{r}_G^t$  
of $\widetilde{\T}_G$. 
\end{lemma}

\vspace{-3mm}
\subsection{Projection to the desired timed runs of $\T_1,\ldots, \T_N$} \label{sec: projection}

An accepting run $\widetilde r_G^t$ of the global B\"uchi WTS $\widetilde \T_G$ can be found efficiently leveraging ideas from automata-based LTL model checking \cite{katoen}. Namely, $\widetilde \T_G$ is viewed as a graph that is searched for a so-called accepting lasso; a cycle containing an accepting state that is reachable from the initial state. Once  $\widetilde r_G^t$ is obtained, Lemmas \ref{eq: lemma_3}, \ref{eq: lemma_2}, and \ref{eq: lemma_1} directly provide guidelines for projection of $\widetilde r_G^t$ onto the individual timed runs of $\T_1,\ldots,\T_N$. In particular, $\widetilde r_G^t$ is projected onto a timed run $r_G^t$ of $\T_G$, which is projected onto timed runs $\widetilde r_1^t,\ldots,\widetilde r_N^t$ of $\widetilde \T_1,\ldots,\T_N$, which are finally projected onto timed runs $ r_1^t,\ldots, r_N^t$ of $\T_1,\ldots, \T_N$, respectively. Such a projection guarantees that $ r_1^t,\ldots, r_N^t$  are a solution to Problem \ref{problem: basic_prob}.


\vspace{-2mm}
\section{Illustrative Example} \label{sec: simulation_results}

\begin{figure}[ht!]
\centering
\begin{tikzpicture}[scale = 0.7] 

\draw[step=1.5, line width=.04cm] (-7.5,-1.5) grid (3,3);

\draw[-latex, draw=green!90, line width = 1.0] (-2.25, 3.30) -- (-2.25,2.50);
\draw[-latex, draw=red!70, line width = 1.0] (-2.25, -1.8) -- (-2.25,-1.05);
\draw [green!90, line width = 1.5] (-2.90, 3.30) -- (-2.25,3.30);
\draw [red!50, line width = 1.0] (-2.90, -1.8) -- (-2.25,-1.8);

\filldraw[fill=green!90, line width=.04cm]  (-2.25,2.30) circle (0.25cm);
\node at (-3.20,3.25) {$R_1$};

\filldraw[fill=red!70, line width=.04cm]  (-2.25,-0.80) circle (0.25cm);
\node at (-3.20,-1.80) {$R_2$};

\filldraw[fill=yellow!90, line width=.04cm] (0.28, 1.67) rectangle +(1.0, 1.0);
\node (agent 1) at (0.75, 2.20) [label=center:\textbf{$A$}] {};

\filldraw[fill=yellow!90, line width=.04cm] (-7.25, -1.2) rectangle +(1.0, 1.0);
\node (agent 1) at (-6.75,-0.7) [label=center:\textbf{$B$}] {};

\filldraw[fill=blue!50, line width=.04cm] (-4.28, 1.7) rectangle +(1.0, 1.0);
\filldraw[fill=red!70, line width=.04cm]  (-3.75, 2.2) circle (0.25cm);
\node at (-3.75, 2.2) {$2$};

\filldraw[fill=blue!50, line width=.04cm] (-5.68, 0.3) rectangle +(1.0, 1.0);
\filldraw[fill=green!90, line width=.04cm]  (-5.15, 0.8) circle (0.25cm);
\node at (-5.15, 0.8) {$1$};

\draw[-latex, draw=green!90, line width = 1.0] (-2.25,2.30) -- (-2.25,0.80);
\draw[-latex, draw=green!90, line width = 1.0] (-2.25,0.80) -- (-3.90,0.80);
\draw[-latex, draw=green!90, line width = 1.0] (-3.90,0.80) -- (-5.1,0.80);
\draw[-latex, draw=green!90, line width = 1.0] (-5.10,0.80) -- (-5.1,-0.40);
\draw[-latex, draw=green!90, line width = 1.0] (-5.10,-0.40) -- (-3.8,-0.40);
\draw[-latex, draw=green!90, line width = 1.0] (-3.8,-0.40) -- (-2.2,-0.40);
\draw[-latex, draw=green!90, line width = 1.0] (-2.2,-0.40) -- (-0.6,-0.40);
\draw[-latex, draw=green!90, line width = 1.0] (-0.6,-0.40) -- (0.7,-0.40);
\draw[-latex, draw=green!90, line width = 1.0] (0.7,-0.40) -- (0.7,0.7);
\draw[-latex, draw=green!90, line width = 1.0] (0.7,0.7) -- (0.7,1.9);

\draw[-latex, draw=red!70, line width = 1.0] (-2.25,-0.80) -- (-3.70,-0.8);
\draw[-latex, draw=red!70, line width = 1.0] (-3.70,-0.8) -- (-3.70,0.7);
\draw[-latex, draw=red!70, line width = 1.0] (-3.70,0.7) -- (-3.70,2.2);
\draw[-latex, draw=red!70, line width = 1.0] (-3.70,2.2) -- (-2.2,2.2);
\draw[-latex, draw=red!70, line width = 1.0] (-2.2,2.2) -- (-0.7,2.2);
\draw[-latex, draw=red!70, line width = 1.0] (-0.7,2.2) -- (0.6,2.2);

\filldraw[fill=blue!50, line width=.04cm] (1.65, 1.6) rectangle +(1.2, 1.2);
\filldraw[fill=green!90, line width=.04cm]  (2.0, 2.1) circle (0.20cm);
\filldraw[fill=red!70, line width=.04cm]  (2.5, 2.1) circle (0.20cm);
\node at (2.0, 2.1) {$1$};
\node at (2.5, 2.1) {$2$};

\node at (-7.2, 2.7) {$\pi_1$};
\node at (1.9, -0.3) {$\pi_{21}$};
\node at (1.9, 1.30) {$\pi_{14}$};
\node at (-7.2, 1.20) {$\pi_{8}$};
\end{tikzpicture}
\caption{An illustrative example with $2$ robots evolving in a common workspace.  
Let $\mathcal{W}_0 = \pi_1 \cup \ldots \cup \pi_{21}$. We enumerate the regions starting from the left region in every row and ending in the right. The initial positions of robots $R_1, R_2$ are depicted by a green and a red circle, respectively, the desired meeting points in yellow and the recharging spots by the agents' respective colors inside a blue box. The accepting runs for task specifications $\phi_1$, $\phi_2$, $\phi_G$ are depicted with green and red arrows for agent 1 and agent 2 respectively.}
\label{fig: illustrative_example}
\end{figure}
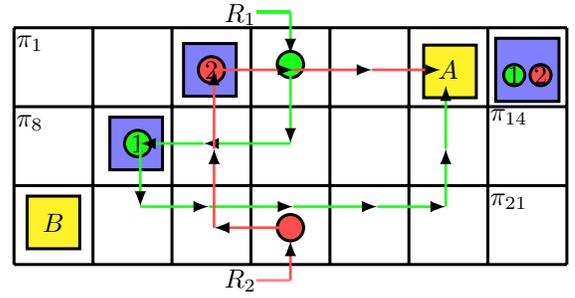

For an illustrative example, consider $2$ robots in the shared workspace of Fig. \ref{fig: illustrative_example}. The workspace is partitioned into $W = 21$ cells and a robot's state is defined by the cell it is currently present at. Agent 1 (R1) is depicted in green and it is two times faster than Agent 2 (R2) which is depicted in red. We assume that the environment imposes such moving constraints that the traveling right and up is faster than left and down. Let Agent 1 need 1 time unit for up and right moves and 2 time units for down and left moves. Let also Agent 2 need 2 time units for up and right moves and  4 time units for down and left moves.

We consider a scenario where the robots have to eventually meet at yellow regions (global team task), and at the same time, they have to recharge within a certain time interval in recharge locations (blue squares with the circles in the respective color). The  individual specifications are $\varphi_1 = \Diamond_{\leq 6} (\mathit{recharge1})$ and $\varphi_2 = \Diamond_{\leq 12} (\mathit{recharge2})$ stating that agent 1 has to recharge within 5 time units and agent 2 within 10 units, respectively, and the team task is $\varphi_G = \Diamond_{\leq 30} \{(\mathit{meet_1^A} \wedge \mathit{meet_2^A}) \vee (\mathit{meet_1^B} \wedge \mathit{meet_2^B} ) \}$ stating that the agents have to meet either in yellow region $A$ or $B$ within 30 time units.

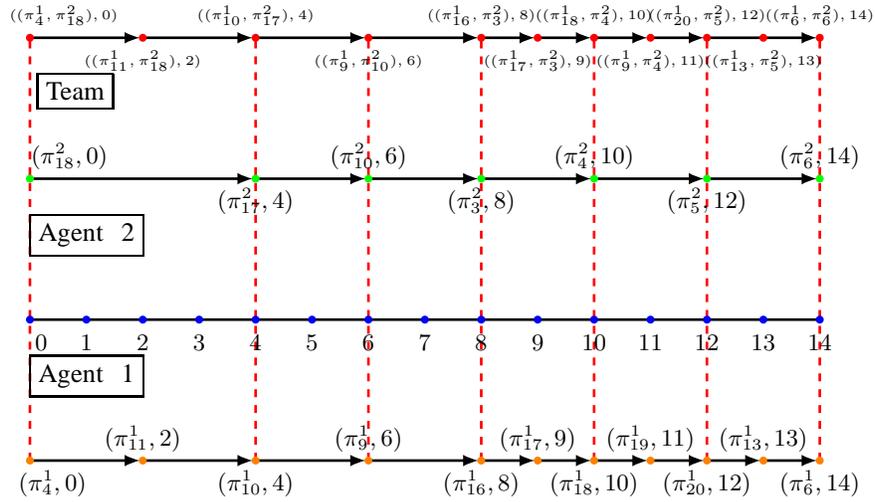
\begin{figure*} [htp!] \centering
\begin{tikzpicture} [scale = 0.75]


\draw[dashed, line width = 1.0, red] (0,-2.5) -- (0,5);
\draw[dashed, line width = 1.0, red] (4,-2.5) -- (4,5);
\draw[dashed, line width = 1.0, red] (6,-2.5) -- (6,5);
\draw[dashed, line width = 1.0, red] (8,-2.5) -- (8,5);
\draw[dashed, line width = 1.0, red] (10,-2.5) -- (10,5);
\draw[dashed, line width = 1.0, red] (12,-2.5) -- (12,5);
\draw[dashed, line width = 1.0, red] (14,-2.5) -- (14,5);

\draw[-latex, line width = 1.0] (0, -2.5) -- (1.95,-2.5);
\draw[-latex, line width = 1.0] (1.95, -2.5) -- (3.95,-2.5);
\draw[-latex, line width = 1.0] (3.95, -2.5) -- (5.95,-2.5);
\draw[-latex, line width = 1.0] (5.95, -2.5) -- (7.95,-2.5);
\draw[-latex, line width = 1.0] (7.95, -2.5) -- (8.95,-2.5);
\draw[-latex, line width = 1.0] (8.95, -2.5) -- (9.95,-2.5);
\draw[-latex, line width = 1.0] (9.95, -2.5) -- (10.95,-2.5);
\draw[-latex, line width = 1.0] (10.95, -2.5) -- (11.95,-2.5);
\draw[-latex, line width = 1.0] (11.95, -2.5) -- (12.95,-2.5);
\draw[-latex, line width = 1.0] (12.95, -2.5) -- (13.95,-2.5);

\draw[-latex, line width = 1.0] (0, 2.5) -- (3.95,2.5);
\draw[-latex, line width = 1.0] (3.95, 2.5) -- (5.95,2.5);
\draw[-latex, line width = 1.0] (5.95, 2.5) -- (7.95,2.5);
\draw[-latex, line width = 1.0] (7.95, 2.5) -- (9.95,2.5);
\draw[-latex, line width = 1.0] (9.95, 2.5) -- (11.95,2.5);
\draw[-latex, line width = 1.0] (11.95, 2.5) -- (13.95,2.5);

\draw[-latex, line width = 1.0] (0, 5) -- (1.95,5);
\draw[-latex, line width = 1.0] (1.95, 5) -- (3.95,5);
\draw[-latex, line width = 1.0] (3.95, 5) -- (5.95,5);
\draw[-latex, line width = 1.0] (5.95, 5) -- (7.95,5);
\draw[-latex, line width = 1.0] (7.95, 5) -- (8.95,5);
\draw[-latex, line width = 1.0] (8.95, 5) -- (9.95,5);
\draw[-latex, line width = 1.0] (9.95, 5) -- (10.95,5);
\draw[-latex, line width = 1.0] (10.95, 5) -- (11.95,5);
\draw[-latex, line width = 1.0] (11.95, 5) -- (13.95,5);

\draw[line width = 1.0] (0,0) -- (14,0);

\fill[blue] (0,0) circle (2pt);
\fill[blue] (1,0) circle (2pt);
\fill[blue] (2,0) circle (2pt);
\fill[blue] (3,0) circle (2pt);
\fill[blue] (4,0) circle (2pt);
\fill[blue] (5,0) circle (2pt);
\fill[blue] (6,0) circle (2pt);
\fill[blue] (7,0) circle (2pt);
\fill[blue] (8,0) circle (2pt);
\fill[blue] (9,0) circle (2pt);
\fill[blue] (10,0) circle (2pt);
\fill[blue] (11,0) circle (2pt);
\fill[blue] (12,0) circle (2pt);
\fill[blue] (13,0) circle (2pt);
\fill[blue] (14,0) circle (2pt);

\fill[orange] (0,-2.5) circle (2pt);
\fill[orange] (2,-2.5) circle (2pt);
\fill[orange] (4,-2.5) circle (2pt);
\fill[orange] (6,-2.5) circle (2pt);
\fill[orange] (8,-2.5) circle (2pt);
\fill[orange] (9,-2.5) circle (2pt);
\fill[orange] (10,-2.5) circle (2pt);
\fill[orange] (11,-2.5) circle (2pt);
\fill[orange] (12,-2.5) circle (2pt);
\fill[orange] (13,-2.5) circle (2pt);
\fill[orange] (14,-2.5) circle (2pt);

\fill[green] (0,2.5) circle (2pt);
\fill[green] (4,2.5) circle (2pt);
\fill[green] (6,2.5) circle (2pt);
\fill[green] (8,2.5) circle (2pt);
\fill[green] (10,2.5) circle (2pt);
\fill[green] (12,2.5) circle (2pt);
\fill[green] (14,2.5) circle (2pt);

\fill[red] (0, 5) circle (2pt);
\fill[red] (2, 5) circle (2pt);
\fill[red] (4, 5) circle (2pt);
\fill[red] (6, 5) circle (2pt);
\fill[red] (8, 5) circle (2pt);
\fill[red] (9, 5) circle (2pt);
\fill[red] (10,5) circle (2pt);
\fill[red] (11,5) circle (2pt);
\fill[red] (12,5) circle (2pt);
\fill[red] (13,5) circle (2pt);
\fill[red] (14,5) circle (2pt);

\node at (0.2, -0.4) {\small $0$};   
\node at (1, -0.4) {\small $1$};      
\node at (2, -0.4) {\small $2$};      
\node at (3, -0.4) {\small $3$};      
\node at (4, -0.4) {\small $4$};      
\node at (5, -0.4) {\small $5$};     
\node at (6, -0.4) {\small $6$};     
\node at (7, -0.4) {\small $7$}; 
\node at (8, -0.4) {\small $8$}; 
\node at (9, -0.4) {\small $9$}; 
\node at (10, -0.4) {\small $10$}; 
\node at (11, -0.4) {\small $11$}; 
\node at (12, -0.4) {\small $12$};
\node at (13, -0.4) {\small $13$}; 
\node at (14, -0.4) {\small $14$};

\node at (0.4, -2.9) {\small $(\pi_4^1, 0)$};  
\node at (2.0, -2.1) {\small $(\pi_{11}^1, 2)$};  
\node at (4, -2.9) {\small $(\pi_{10}^1, 4)$};  
\node at (6, -2.1) {\small $(\pi_{9}^1, 6)$};  
\node at (8, -2.9) {\small $(\pi_{16}^1, 8)$};  
\node at (9, -2.1) {\small $(\pi_{17}^1, 9)$};  
\node at (10, -2.9) {\small $(\pi_{18}^1, 10)$};  
\node at (11, -2.1) {\small $(\pi_{19}^1, 11)$};  
\node at (12, -2.9) {\small $(\pi_{20}^1, 12)$};
\node at (13, -2.1) {\small $(\pi_{13}^1, 13)$};   
\node at (14, -2.9) {\small $(\pi_{6}^1, 14)$};


\node at (0.7, 2.9) {\small $(\pi_{18}^2, 0)$};
\node at (4, 2.1) {\small $ (\pi_{17}^2, 4)$};
\node at (6, 2.9) {\small $(\pi_{10}^2, 6)$};
\node at (8, 2.1) {\small $(\pi_{3}^2, 8)$};
\node at (10, 2.9) {\small $(\pi_{4}^2, 10)$};
\node at (12, 2.1) {\small $(\pi_{5}^2, 12)$};
\node at (14, 2.9) {\small $(\pi_{6}^2, 14)$};

\node at (0.6, 5.4) {\tiny $((\pi_{4}^1, \pi_{18}^2), 0)$};
\node at (2, 4.6) {\tiny $((\pi_{11}^1, \pi_{18}^2), 2)$};
\node at (4, 5.4) {\tiny $((\pi_{10}^1, \pi_{17}^2), 4)$};
\node at (6, 4.6) {\tiny $((\pi_{9}^1, \pi_{10}^2), 6)$};
\node at (8, 5.4) {\tiny $((\pi_{16}^1, \pi_{3}^2), 8)$};
\node at (9, 4.6) {\tiny $((\pi_{17}^1, \pi_{3}^2), 9)$};
\node at (10, 5.4) {\tiny $((\pi_{18}^1, \pi_{4}^2), 10)$};
\node at (11, 4.6) {\tiny $((\pi_{9}^1, \pi_{4}^2), 11)$};
\node at (12, 5.4) {\tiny $((\pi_{20}^1, \pi_{5}^2), 12)$};
\node at (13, 4.6) {\tiny $((\pi_{13}^1, \pi_{5}^2), 13)$};
\node at (14, 5.4) {\tiny $((\pi_{6}^1, \pi_{6}^2), 14)$};

\node at (1.0, 1.5) {\boxed{$\text{Agent} \ 2$}};
\node at (1.0, -1.0) {\boxed{$\text{Agent} \ 1$}};
\node at (0.8, 4.05) {\boxed{$\text{Team}$}};
\end{tikzpicture}
\caption{The accepting runs $\widetilde{r}_1^t, \widetilde{r}_2^t$, the collective run $\widetilde{r}_G^t$ and the corresponding timed stamps. We denote with red dashed lines the times that both agents have the same time stamps}
\label{fig: run_robots2}
\end{figure*}

By following the process that was described in Section \ref{sec: solution} step by step we have that an accepting timed run is
$\widetilde{r}_G^t = ((\pi_4^1, \pi_{18}^2), 0)((\pi_{11}^1, \pi_{18}^2), 2) \ldots ((\pi_9^1,\pi_{10}^2), 6) ((\pi_{16}^1,\pi_{3}^2), 8) \ldots \\ ((\pi_{13}^1, \pi_{5}^2), 13)((\pi_6^1, \pi_6^2), 14) \ldots$ with corresponding timed word $w(\widetilde{r}_G^t) = (\emptyset, 0)(\emptyset, \pi_{18}^2), 2) \ldots (\{\mathit{recharge1}\}, 6) \\  (\{\mathit{recharge2}\}, 8) \ldots (\emptyset, \pi_{5}^2), 13)((\{\mathit{meet}_1^A, \mathit{meet}_2^A \}, 14) \ldots$ which satisfies formula $\phi_G$. The run $\widetilde{r}_G^t$ can be projected onto individual the timed runs $\widetilde{r_1}^t = (\pi_4^1, 0)(\pi_{11}^1, 2)(\pi_{10}^1, 4) \\ (\pi_{9}^1, 6)(\pi_{16}^1, 8)(\pi_{17}^1, 9)(\pi_{18}^1, 10)(\pi_{19}^1, 11)(\pi_{20}^1, 12)(\pi_{13}^1, 13) \\ 
(\pi_{6}^1, 14) \ldots$ and $\widetilde{r_2}^t = (\pi_{18}^2, 0)(\pi_{17}^2, 4)(\pi_{10}^2, 6)(\pi_{3}^2, 8)(\pi_{4}^2, 10)\\(\pi_{5}^2, 12)(\pi_{6}^2, 14) \ldots$ (they are depicted in Fig. \ref{fig: illustrative_example} with green and red arrows respectively) with corresponding timed words $w(\widetilde{r}_1^t) = (\emptyset, 0)(\emptyset, 2)(\emptyset, 4)(\{\mathit{recharge1}\}, 6)(\emptyset, 8)(\emptyset, 9)(\emptyset, 10)(\emptyset, 11)\\(\emptyset, 12)(\emptyset, 13)(\{\mathit{meet}_1^A\}, 14) \ldots$ and $w(r_2^t) = (\emptyset, 0)(\emptyset, 4)\\(\emptyset, 6)(\{\mathit{recharge2}\}, 8)(\emptyset, 10)(\emptyset, 12)(\{\mathit{meet}_2^A\}, 14)\ldots$ which satisfy formulas $\phi_1$ and $\phi_2$ respectively. All conditions from \eqref{eq: problem_adf} are satisfied. The runs and the words of the illustrative example are depicted in Fig. \ref{fig: run_robots2}.

Consider now the alternative runs of the agents, where they first meet in the meeting point $A$ (after 8 time units) and then recharge in the region $\pi_{7}$ (after 9 and 10 time units, respectively). Regardless of how the agents continue, they have accomplished the untimed formulas $\varphi_1' = \Diamond(\mathit{recharge1})$, $\varphi_2' = \Diamond (\mathit{recharge2})$, and $\varphi_G' = \Diamond \{(\mathit{meet_1^A} \wedge \mathit{meet_2^A}) \vee (\mathit{meet_1^B} \wedge \mathit{meet_2^B} ) \}$. Although this is in fact a more efficient way to satisfy the untimed formulas $\varphi_1', \varphi_2'$, and $\varphi_G'$ than the one described above, the formula $\varphi_1$ is violated due to its time constraint.

%
\vspace{-3mm}
\section{Conclusions and Future Work} \label{sec: conclusions}
We have proposed a systematic method for multi-agent controller synthesis aiming cooperative planning under high-level specifications given in MITL formulas. The solution involves a sequence of algorithmic automata constructions such that not only team specifications but also individual specifications should be fulfilled. Future research directions include the consideration of more complicated dynamics than the fully actuated ones in \eqref{eq: system}, the decentralized solution such that every agent has information only from his neighbors as well as the modeling of the system with Markov Decision Processes (MDPs) and probabilistic verification.





\vspace{-1mm}

\bibliography{bibfile}
\bibliographystyle{ieeetr}
\addtolength{\textheight}{-12cm}

\end{document}